\documentclass[prb,showpacs,preprint]{revtex4}
\usepackage{graphics}

\begin{document}

\title{\bf  Modification of the Charge ordering in 
Pr$_{1/2}$Sr$_{1/2}$MnO$_{3}$ Nanoparticles} 

\author{Anis Biswas, I. Das}
\author{Chandan Majumdar}

\affiliation{ Saha Institute of Nuclear Physics,1/AF,Bidhannangar,
Kolkata 7000 064, India}

%%%%%%%%%%%%%%%%%%%%%%%%%%%%%%%%%%%%%%%%%%%%%%%%%%%%%%%%%%%%%%%%%%%%%%%%%%
\begin{abstract} 
Transport and magnetic properties have been studied in two sets of
 sol-gel prepared
 Pr$_{1/2}$Sr$_{1/2}$MnO$_{3}$ nanoparticles having average particle size of
 $30$ nm and $45$ nm. Our measurements suggest that the formation of
 charge ordered state is largely
 affected due to lowering of
 particle size, but the ferromagnetic transition temperature ($T_{C}$) remains unaffected.

\end{abstract}
%%%%%%%%%%%%%%%%%%%%%%%%%%%%%%%%%%%%%%%%%%%%%%%%%%%%%%%%%%%%%%%%%%%%%%%%%%
\maketitle
%%%%%%%%%%%%%%%%%%%%%%%%%%%%%%%%%%%%%%%%%%%%%%%%%%%%%%%%%%%%%%%%%%%%%%%%%%
\section{Introduction} 
\noindent
Interests have grown recently towards the study of hole doped 
Perovskite Manganites with general formula R$_{1-x}$A$_{x}$MnO$_{3}$
(R - Trivalent Rare earth, A - Bivalent ions) because they exhibit intriguing
 phenomena like Colossal Magnetoresistance (CMR), Charge ordering (CO) etc. 
\cite{physics,collosal}. Charge ordering is observed for commensurate fraction 
of concentration of bivalent ion such as $1/8$, $1/3$ and $1/2$ 
\cite{physics,collosal}. This phenomenon draws considerable attention because 
it exhibits a close interrelation among magnetic ordering, electronic transport and crystal structure. Large magnetoresistance (MR) also originates due to
 the melting of CO state by magnetic field \cite{ rev}.
 There are several reports on this phenomena in the 
$x \sim 1/2$ systems like Nd$_{1-x}$Sr$_{x}$MnO$_{3}$ \cite{kuwahara},
 La$_{1-x}$Ca$_{x}$MnO$_{3}$ \cite{wollan,schiffer}, Pr$_{1-x}$Ca$_{x}$MnO$_{3}$ \cite{jirak,pollert,tomioka,yoshizawa},
 Pr$_{1-x}$Sr$_{x}$MnO$_{3}$ \cite{knizek,tomioka2,kajimoto}.
 Most of the above mentioned materials exhibit C-E type  charge ordering. 
However, very few of them with relatively large single electron 
band-width show stripe like A-type antiferromagnetic charge ordering. As the 
A-type antiferromagnets are made of alternately stacked ferromagnetic planes, 
the ferromagneism and antiferromagnetism are closely connected in such systems.
Therefore, the materials exhibiting A-type antiferromagnetic charge ordering 
offer an opportunity to study the competition of ferromagnetic double exchange
interaction and antiferromagnetic super exchange interaction in charge ordered
lattice.     
In case of  single crystalline form of Pr$_{1/2}$Sr$_{1/2}$MnO$_{3}$ two 
transitions occur.  During cooling from $300$ K it undergoes paramagnetic to 
ferromagnetic state at $\sim$ $270$ K and then ferromagnetic to charge ordered
antiferromagnetic state at $\sim$ $140$ K \cite{tomioka2}.  Insulator to metal 
transition also occurs in  coincidence with the paramagnetic to ferromagnetic
transition at $\sim$ $270$ K and resistivity starts to increase
sharply with decreasing temperature below $\sim$ $140$ K due to the charge
order transition \cite{tomioka2}. Recently, through neutron diffraction study 
it has been conclusively proved that in this sample the charge ordering is not
conventional checker board type (C-E type); it is novel stripe like A-type 
antiferromagnetic \cite{kajimoto} . The detection of this novel kind of charge 
ordering may not  be straightforward and it  has not been clearly detected in 
earlier neutron diffraction study \cite{kawano}. Though there are considerable 
studies in the single crystalline and polycrystalline bulk form of 
Pr$_{1/2}$Sr$_{1/2}$MnO$_{3}$ regarding charge ordering, but this phenomena is 
not explored much in its nanocrystalline form. In this report, our primary 
objective is to study the modification of the charge ordering in case of the 
nanoparticles of Pr$_{1/2}$Sr$_{1/2}$MnO$_{3}$. With this motivation, the 
nanoparticles of Pr$_{1/2}$Sr$_{1/2}$MnO$_{3}$ have been synthesized by sol-gel 
method. Since charge ordering in single crystalline form of 
Pr$_{1/2}$Sr$_{1/2}$MnO$_{3}$ has strong  signature in transport and magnetic 
properties, we have performed transport, magnetotransport and magnetization 
measurements in details to study the charge ordering in case of the 
above mentioned nanoparticles.  From our measurements, it appears that the 
formation of CO state is hindered due to the lowering of grain size and the 
transition from ferromagnetic state to antiferromagnetic CO state is not clearly evident. However, the ferromagnetic transition temperature remains unaffected.

%%%%%%%%%%%%%%%%%%%%%%%%%%%%%%%%%%%%%%%%%%%%%%%%%%%%%%%%%%%%%%%%%%%%%%%%%%%%%%
\section{Sample Preparation and Characterization}
Nanoparticles of Pr$_{1/2}$Sr$_{1/2}$MnO$_{3}$ have been  prepared by sol-gel 
method. The starting materials are Pr$_{6}$O$_{11}$, MnO$_{2}$ and SrNO$_{3}$ 
with the purity of $99.9$ \%. The oxides have been converted to nitrates and 
suitable amount of citric acid is mixed with the water solution of the nitrates.Then the solution has been slowly evaporated until a gel is 
formed. The gel is  decomposed into a black porous powder.
This porous powder is then divided into several parts, which are kept under 
heat treatment at different temperatures
to prepare samples of  different particle sizes. Powder X-ray diffraction 
pattern (XRD) shows that single
phased sample is formed  when sample is heated at temperature $1200^{o}$ C and 
above. Two sets of sample of different
particle sizes have been prepared by  heat- treatment for $4$ hours at 
$1200^{o}$ C and $1400^{o}$ C respectively.
From the Transmission Electron Microscopy  measurements,  average particle 
sizes are determined as $30$ nm
and $45$ nm for $1200^{o}$ C and $1400^{o}$ C heat treated samples.
To compare the results of nanocrystalline samples with polycrystalline
bulk sample, we have also prepared polycrystalline bulk form of 
Pr$_{1/2}$Sr$_{1/2}$MnO$_{3}$ by solid state reaction from the stoichiometric
ratio of Pr$_{6}$O$_{11}$, MnO$_{2}$ and SrCO$_{3}$. In this case, after several
 intermediate heat treatments at $1000^{o}$C, final heat treatment has been 
given at $1500^{o}$ C for $22$ hours. The XRD pattern confirms the
 single phase nature of the sample
in this case also.
%%%%%%%%%%%%%%%%%%%%%%%%%%%%%%%%%%%%%%%%%%%%%%%%%%%%%%%%%%%%%%%%%%%%%%%%%%%%%%%

\section{Experimental Results and Discussion} 
The resistivity measurements have been performed in usual four probe method.
The temperature dependence of resistivity in the temperature range 
$3.3$ K - $300$ K for the polycrystalline bulk 
Pr$_{1/2}$Sr$_{1/2}$MnO$_{3}$ sample (Fig. 1) shows that there is an insulator 
to metal transition (I-M) at around $275$ K. As temperature is 
decreased further, resistivity starts to increase sharply below $\sim$  $130$ K. These two temperatures are in good agreement with the reported temperatures 
where I-M transition (paramagnetic to ferromagnetic) and the charge order 
transition are observed for single crystalline form of the sample 
\cite{tomioka2}. The sharp increase of resistivity with temperature below 
$\sim$ $130$ K is believed to be due to charge order transition.

A commercial SQUID magnetometer has been employed for magnetization 
measurements. The zero field cooled (ZFC) DC susceptibility with temperature 
has been studied in 
presence of $20$ kOe magnetic field. The temperature dependence of DC
susceptibility for the polycrystalline bulk sample indicates that the 
paramagnetic to ferromagnetic transition occurs at $\sim$ $275$ K where the 
I-M transition in resistivity measurement is observed. There is 
also a ferromagnetic to antiferromagnetic transition at around $115$ K (Fig. 2) which is slightly lower than the temperature at which resistivity starts to 
increase sharply with the lowering of temperature (Fig. 1). The transition 
temperatures are assigned as the temperatures where the fastest change of 
magnetization with temperature takes place. For the bulk sample, the magnetic 
field dependence of magnetization is linear at $350$ K indicating paramagnetic 
behavior. The ferromagnetic state in the temperature range between  $275$ K and $115$ K has been confirmed by M(H) measurements at $225$ K (Fig. 3).The 
M(H) measurements at $3.3$ K also confirms the antiferromagnetic state below 
$115$ K (Fig. 3). For the bulk sample, an abrupt increase of magnetization 
occurs at 
around $55$ kOe in M(H) data at $3.3$ K (Fig. 3). It can be attributed to the 
metamagnetic transition due to melting of CO state by magnetic field. An 
appreciable hysteresis is also observed 
in M(H). The polycrystalline bulk sample shows all the  features
and transition temperatures of Pr$_{1/2}$Sr$_{1/2}$MnO$_{3}$  
which are consistent with the literature \cite{tomioka2} and the results of the sample can be used to compare with the results obtained for nanocrystalline 
samples.    

The study of variation of resistivity with temperature for nanoparticles of 
different sizes (Fig. 1) reveals that there is no indication of I-M transition 
in the entire temperature range ($3.3$ - $300$ K). Instead of this resistivity 
increases with the decrease of temperature down to $3.3$ K. This increase is 
reasonably fast in the low temperature region. But the increase of resistivity
 is not as sharp as in the case of polycrystalline sample.
 From the temperature dependence of resistivity for nanocrystalline samples, it is quite 
difficult  to comment whether the increase of resistivity are connected 
with charge ordering or some other effect.

The temperature dependence of ZFC susceptibility for nanoparticles shows a 
transition from paramagnetic to ferromagnetic state at around $275$ K which is 
also $T_{C}$ for bulk sample (Fig. 2). However, ferromagnetic to 
antiferromagnetic transition is not evident for both the nanocrystalline 
samples. The ZFC susceptibility measurement has been carried out in the presence
of $20$ kOe magnetic field. For both the samples, M(H) behavior at $350$ K is 
linear confirming the paramagnetic state of the samples.
There is a decrease in the rate of change of magnetization from temperature 
around $140$ K in comparison to the higher temperature region. It appears that 
 the value of susceptibility for $45$ nm particle sized sample is larger than 
that for other nanocrystalline sample  in temperature region above 
$\sim$ $140$ K and below this temperature susceptibility of $30$ nm sample is 
larger than that of $45$ nm sample. 
We further study the variation of field cooled (FC) and zero field cooled 
susceptibility with temperature in the presence of $100$ Oe magnetic field for
 the nanocrystalline samples. The FC and ZFC susceptibility curve for
the sample of particle size $45$ nm has been shown in Fig. 4. A large 
bifurcation between ZFC and FC susceptibility has been observed below 
$T_{C}$ . 
Usually nanoparticles can be considered to be composed of two parts. The inner 
part is core and the outer part is grain boundary or surface layer. The surface layers are in disordered magnetic state comprising of non-collinear spin arrangement as well as defects,vacancies, stress and broken bonds 
\cite{kodama,crisan,coey,zhi}.  Due to the existence of such a magnetically 
disordered surface layer, magnetic frustration can occur. As a result 
a cluster glass state may form which causes the bifurcation between FC and 
ZFC susceptibility curve below $T_{C}$. The origins of the finer features
in FC and ZFC curves are not clear at present.  
       
The magnetic field dependence of magnetization for nanoparticles shows 
ferromagnetic nature even at the lowest temperature $3.3$ K (Fig. 3). For both 
the samples at $225$ K and $3.3$ K, magnetization  does not saturate even 
at $70$ kOe magnetic field (Fig. 3). This type of unsaturated magnetization
even in presence of quite high magnetic field for nanoparticles has been
observed in some earlier experimental studies \cite{crisan,david}. 
Magnetization for $45$ nm sample is larger than that in $30$ nm for all field 
values at $225$ K. However, at $3.3$ K the situation reverses, there 
magnetization for $30$ nm sample is larger than that for $45$ nm.

From the magnetization measurements on nanocrystalline samples, no clear 
evidence of ferromagnetic to  antiferromagnetic transition has been observed,
but the ferromagnetic transition temperature remains almost same as that of 
bulk sample.  As the particle size is reduced, the formation
of CO state is highly disturbed. The fraction of antiferromagnetic CO 
state in the nanoparticles may be too small to dominate over
the ferromagnetic interaction. As a result, the distinct ferromagnetic to 
antiferromagnetic transition  disappears.

In case of manganite nanoparticle system, the ferromagnetic interaction is 
governed by the core spins inside the grains. Though the reduction of particle
size does not substantially affect the ferromagnetic interaction in case of the nanocrystalline samples, but the existence of the random non-collinear
spins as well as the defects, dislocations etc. in the surface 
layers can reduce the value of magnetization in comparison with the bulk. 
Magnetic field as high as $70$ kOe seems to be not enough to align the surface 
spins. As a result,M (H) (at $225$ K and $3.3$ K) does not saturate (Fig. 3).
For the bigger sized particles the surface effect is less than that for 
the smaller sized particles. Therefore, the susceptibility for $45$ nm 
particle sized sample is larger than that of $30$ nm particle sized sample in
the temperature range above $\sim$ $140$ K (Fig. 2). For the same reason, the
value of magnetization in M(H) at $225$ K for $45$ nm particles is also larger
in comparison with $30$ nm particles for all magnetic field value.      
But in low temperature it appears that the fraction of antiferromagnetic CO 
state in $45$ nm particles slightly dominates over that for $30$ nm particles. 
This is reflected in higher susceptibility value of $30$ nm particles in 
comparison to $45$ nm in temperature below $\sim$ $140$ K  (Fig. 2) and also in M(H) curve at $3.3$ K, where M for $30$ nm sample is larger than 
$45$ nm sample for all magnetic field values (Fig. 3).   
 
The melting of CO state by magnetic field can give rise to large negative
MR. The existence of the non-collinear spin arrangements in  surface layers can 
also play vital role in the magnetotransport phenomena. The temperature 
dependence of resistivity has been studied in the presence of $80$ kOe magnetic 
field for the bulk as well as nanocrystalline samples . Magnetoresistance is 
defined as $[R(H)-R(0)]/R(H)$, where R(H) is resistance in presence of magnetic
field and R(0) is resistance in absence of magnetic field.
For the polycrystalline bulk sample, the variation of negative MR with
temperature indicates that there exist a small peak in negative MR at
around I-M transition temperature ($\sim$ 270 K) (Fig. 5). The negative MR 
starts to increase abruptly with decreasing temperature below $\sim$ 130 K 
which almost coincides with the temperature where zero field resistivity starts
to increase sharply with the lowering of temperature as well as
the reported charge order transition temperature for the bulk sample
\cite{tomioka2} . It reaches $\sim$ $3000$ \% at $3.3$ K. This sharp increase
of negative MR is due to the melting of CO state by magnetic field.
Here we have used the inflationary  definition of MR to clearly indicate the
increase of MR with decreasing temperature up to the lowest temperature.
The negative MR also increases with the decrease of 
temperature in the low temperature for both the nanocrystalline samples 
(Fig. 5). 
The increase of negative MR is quite gradual. The value of MR at 
lowest temperature is substantially lower than that obtained for the bulk 
sample. If we compare the value of negative MR between the two nanocrystalline 
samples, it appears that the value of MR for the  $30$ nm particles is larger 
than that for the other nanoparticle sample above $\sim$ $40$ K. Below that
temperature the value of MR for bigger sized nanocrystalline sample ($45$ nm
particle sized sample) dominates over that for the smaller sized
nanocrystalline sample. The effect of surface layers can also influence the
high field magnetoresistance \cite{ziese,balcells}. The smaller sized particles 
having relatively larger surface effect can exhibit larger high field 
magnetoresistance
than the bigger sized particles for the temperatures above $\sim$ $40$ K.
It seems that the melting of CO fraction by magnetic field in nanoparticles
is the major contributor to MR below $\sim$ $40$ K. As, the fraction of CO 
state is larger in bigger sized particles, the value of negative MR is larger 
for the sample in comparison with
the smaller particle sized sample below that temperature.
  The  difference in the  fraction of charge ordered antiferromagnetic states  
in case of two different particle sized samples at low temperature is expected 
to influence the magnetic field dependence of the magnetoresistance. With this 
motivation resistivity has been measured as a 
function of magnetic field at $3.3$ K for the two nanocrystalline samples.
 To highlight the low field magnetoresistance more clearly we have used
conservative definition of MR  as, $[R(H)-R(0)]/R(0)$ (\%). The variation of MR 
with magnetic field
 ($0-70$ kOe) at $3.3$ K has been shown in Fig. 6.
For both the nanocrystalline samples, there exists  faster
increase of the value of negative MR at low magnetic field which points out
the existence of low field magnetoresistance (LFMR). LFMR, quantified as the 
value of MR by extrapolating the higher field MR to zero field, is observed to 
be $\sim$ $10\%$ for $30$ nm particle. This is larger than that for $45$ nm
particle for which LFMR is $\sim$ $4\%$.
LFMR may originate from  domain wall scattering. It has been discussed earlier
that the fraction of CO state appears to be larger in $45$ nm in comparison to 
$30$ nm 
particle at $3.3$ K. Thus, in case of $30$ nm particle the fraction of 
ferromagnetic state is larger as indicated in fig. 3 [B]. Probably, for this 
reason,  LFMR in this case is also larger than that of $45$ nm particle.
In higher magnetic field region (H $>$ $30$ kOe), the value of MR is higher
 in  $45$ nm in comparison to $30$ nm particle. The high field MR can be
 dominated by the melting of CO state. 
As the fraction of CO state is larger in $45$ nm sample, the value of MR for 
this sample exceeds the value for $30$ nm at high magnetic field.

Though for polycrystalline bulk and nanocrystalline samples, the ferromagnetic 
ordering temperatures are almost same, metallic state is only achieved 
in  polycrystalline bulk sample (Fig. 1). The fraction of CO state seems
 to be too small to give distinct
antiferromagnetic signature in magnetization. However, resistivity increases
 with the  decrease of temperatures as expected for CO sample.
 To get the better understanding of transport property,
we have performed I-V measurements at different temperatures (Fig. 7).
 The I-V characteristics are non-linear in nature for all
temperatures up to $300$ K in case of nanocrystalline samples (Fig. 7). 
It is reported that for the charge ordered samples, the nonlinearity, 
hysteresis as well as negative differential resistance (NDR) in I-V can be 
observed due to the melting of CO state by electric field \cite{guha}. But we 
have not observed hysteresis and NDR in I-V characteristics.
In present case nonlinear behavior in I-V is appeared  even at $300$ K. 
In case of nanocrystalline materials due to crystalline defects in the grain 
boundary regions, potential energy barrier may be developed. It will affect the conduction between grain
to grain. Electrons have to tunnel through the potential barrier. Due to this
 tunneling phenomena,
I-V characteristics can be non linear in nature. In our  nanoparticle system,
 this strong grain boundary effect plays the key role in transport and causes
 the suppression of metallic state
 as well as uprise in resistivity with the decrease of temperature.

%%%%%%%%%%%%%%%%%%%%%%%%%%%%%%%%%%%%%%%%%%%%%%%%%%%%%%%%%%%%%%%%%%%%%%%%%%%%%%% 

\section{Summary}
To summarize, we have shown that the reduction of grain size  largely affects
 the formation of CO state in case of 
Pr$_{1/2}$Sr$_{1/2}$MnO$_{3}$ nanoparticles and the distinct ferromagnetic to
antiferromagnetic transition associated with CO transition is not evident. 
When the grain size is reduced,
$T_{C}$ does not change appreciably in comparison with the bulk sample but
 the metallic state disappears and resistivity increases with the lowering of 
temperature.Through I-V characteristic study, we have argued that this is due to the 
grain boundary effect in the nanoparticle systems. From the magnetization 
studies, it appears that the nanocrystalline samples exhibit  cluster
glass like behavior. The magnetization measurements along with the 
magnetoresistance measurements also reveal that the small CO fraction may be 
present at 
low temperature only and this fraction is larger in $45$ nm particles than that of $30$ nm particles.

\newpage
\noindent
Fig.1. Resistivity as a function of temperature for $30$ nm, $45$ nm
nanocrystalline and polycrystalline bulk sample of 
Pr$_{1/2}$Sr$_{1/2}$MnO$_{3}$. Inset:
temperature dependence of resistivity in high temperature region. I-M
transition temperature (T$_{C}$) and the temperature at which resistivity 
starts to increase sharply (T$_{CO}$) for the bulk sample are indicated by 
arrows. \\
Fig.2. Zero Field Cooled Susceptibility as a function of temperature 
for $30$ nm, $45$ nm nanocrystalline and polycrystalline bulk samples 
of Pr$_{1/2}$Sr$_{1/2}$MnO$_{3}$. Measurement has been done in presence 
of $20$ kOe magnetic field. \\
Fig.3. Magnetization as a function of magnetic field for $30$ nm,
$45$ nm and polycrystalline bulk sample at (A) $225$ K and (B) $3.3$
K. Inset: Magnetization as a function of field in low magnetic field region. \\
Fig.4. Zero Field Cooled and Field Cooled Susceptibility as function of
temperature for the nanocrystalline sample of particle size $45$ nm.
Measurement has been performed at $100$ Oe magnetic field.\\ 
%Fig.5. Magnetization as a function of magnetic field for $30$ nm, 
%$45$ nm and polycrystalline bulk sample at (A) $225$ K and (B) $3.3$  
%K. Inset: Magnetization as a function of field in low magnetic field region. \\
Fig.5. Magnetoresistance {[R(H)-R(0)]/R(H)} as a function of temperature for 
$80$ kOe magnetic field. The sharp increase of negative magnetoresistance with 
decreasing temperature near T$_{CO}$ for the bulk sample is indicated by arrow.  Inset: peak in negative magnetoresistance
around the ferromagnetic transition temperature for the bulk sample.\\
Fig.6. Magnetoresistance as a function of magnetic field for $30 nm$ and
$45 nm$ sized nanocrystalline samples at $3.3$ K. Magnetoresistance
is defined as [R(H)-R(0)]/R(0). Inset: Indication of low field magnetoresistance. \\ 
Fig.7. I-V Characteristics for nanoparticles: (A) $30$ nm 
(B) $45$ nm at $100$ K, $200$ K and $300$ K.

\newpage
\begin{figure}[ht]
\centering {\resizebox{8.0cm}{6.5cm}{\includegraphics{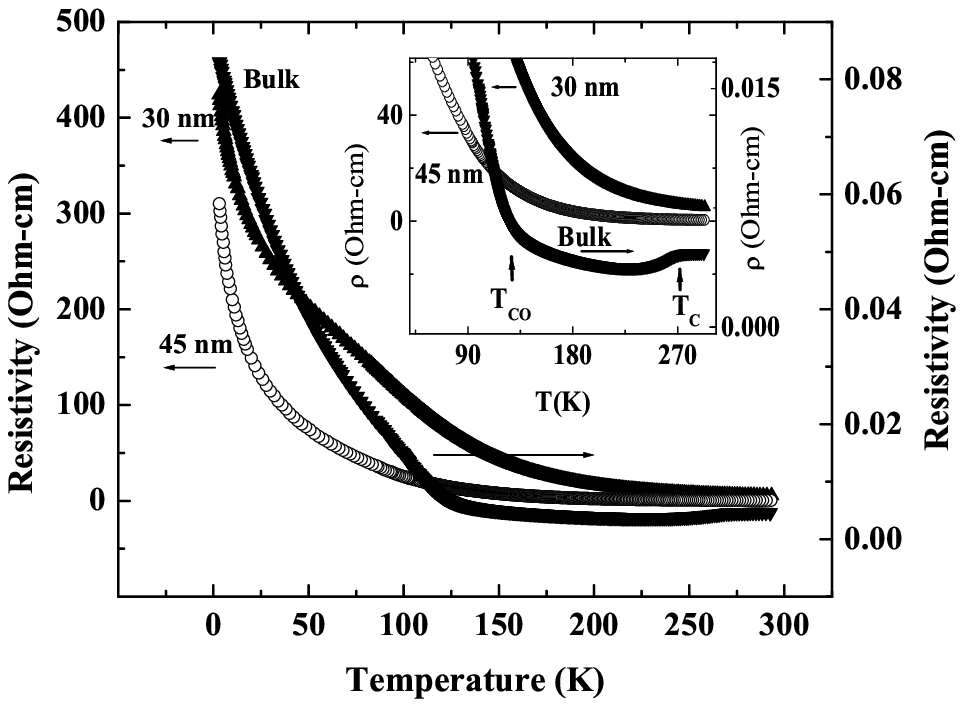}}}
\caption{}
\end{figure}

\newpage
\begin{figure}[ht]
\centering {\resizebox{8.0cm}{6.5cm}{\includegraphics{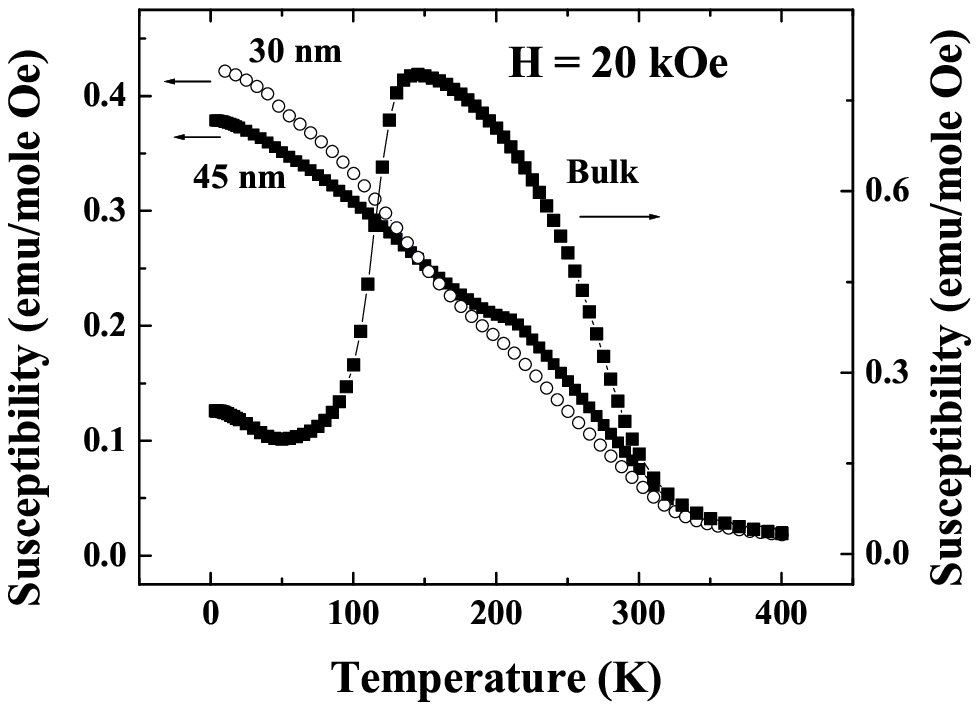}}}
\caption{}
\end{figure}

\newpage
\begin{figure}[ht]
\centering {\resizebox{8cm}{7.5cm}{\includegraphics{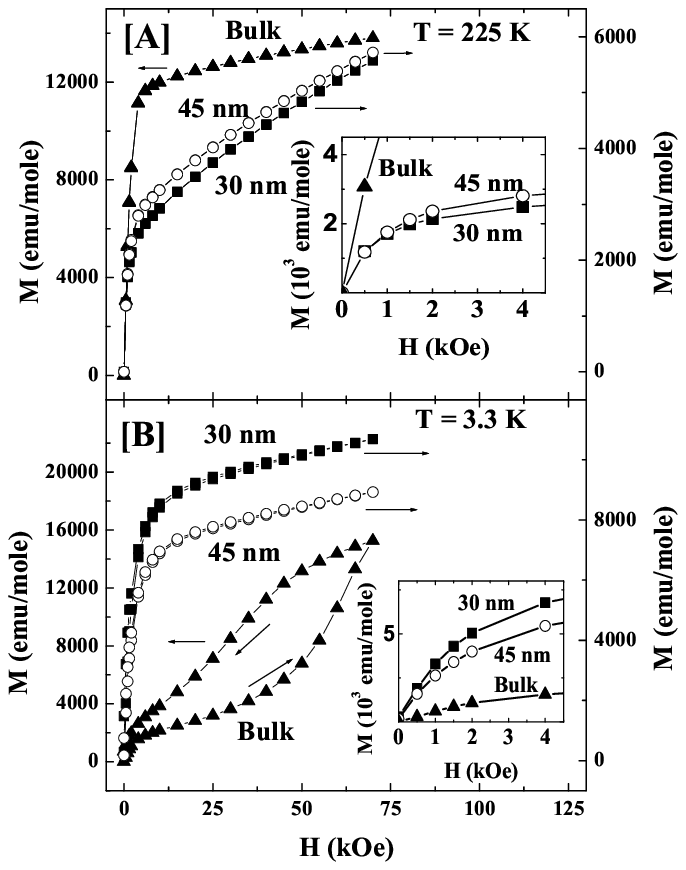}}}
\caption{}
\end{figure}

\newpage
\begin{figure}[ht]
\centering {\resizebox{8cm}{6.5cm}{\includegraphics{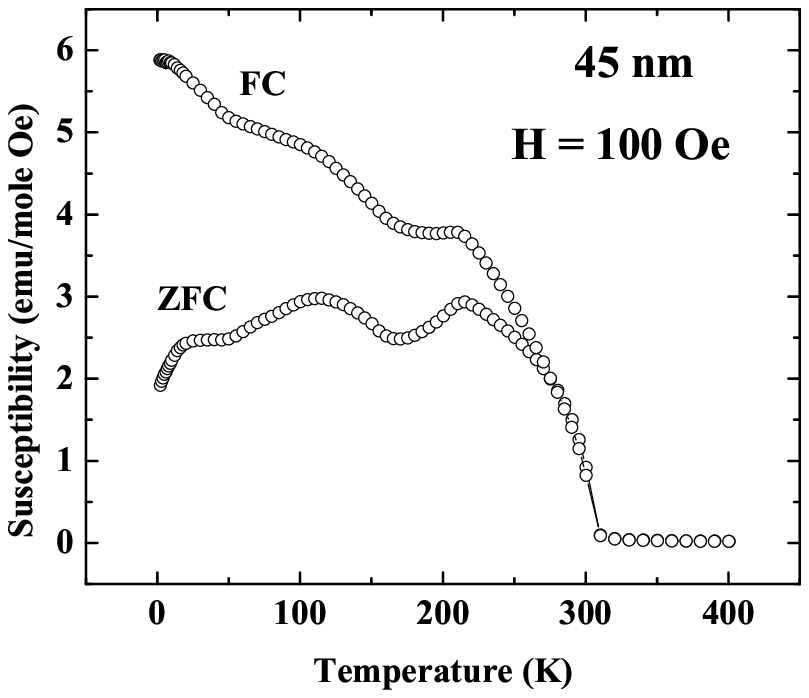}}}
\caption{}
\end{figure}

\newpage
\begin{figure}[ht]
\centering {\resizebox{8.0cm}{6.5cm}{\includegraphics{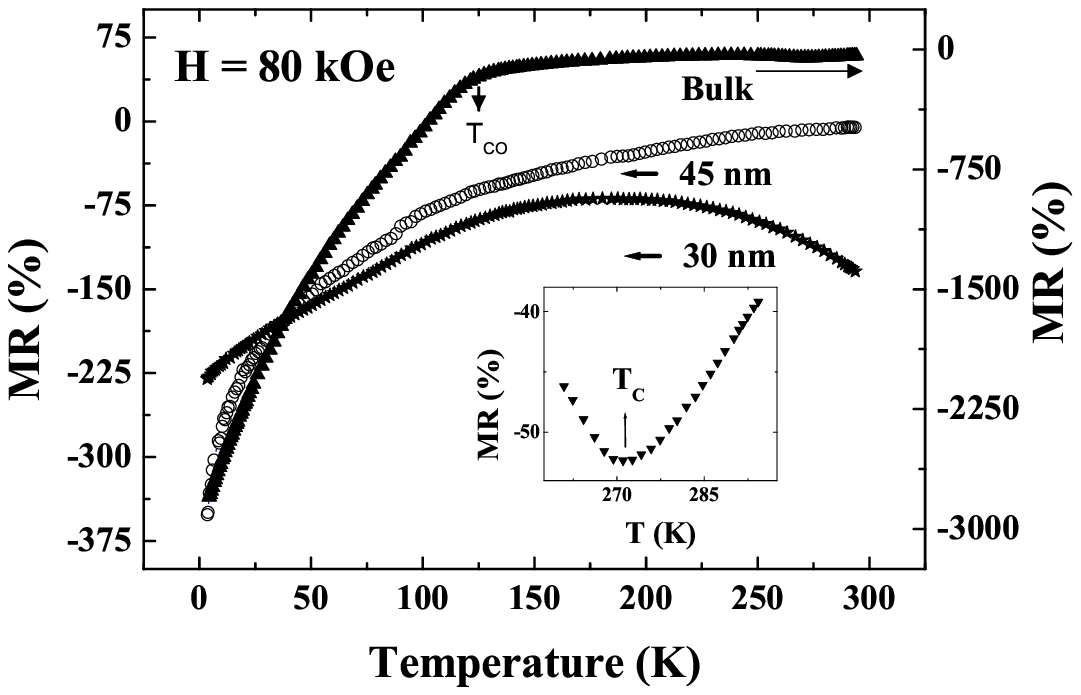}}}
\caption{}
\end{figure}

\newpage
\begin{figure}[ht]
\centering {\resizebox{8.0cm}{6.5cm}{\includegraphics{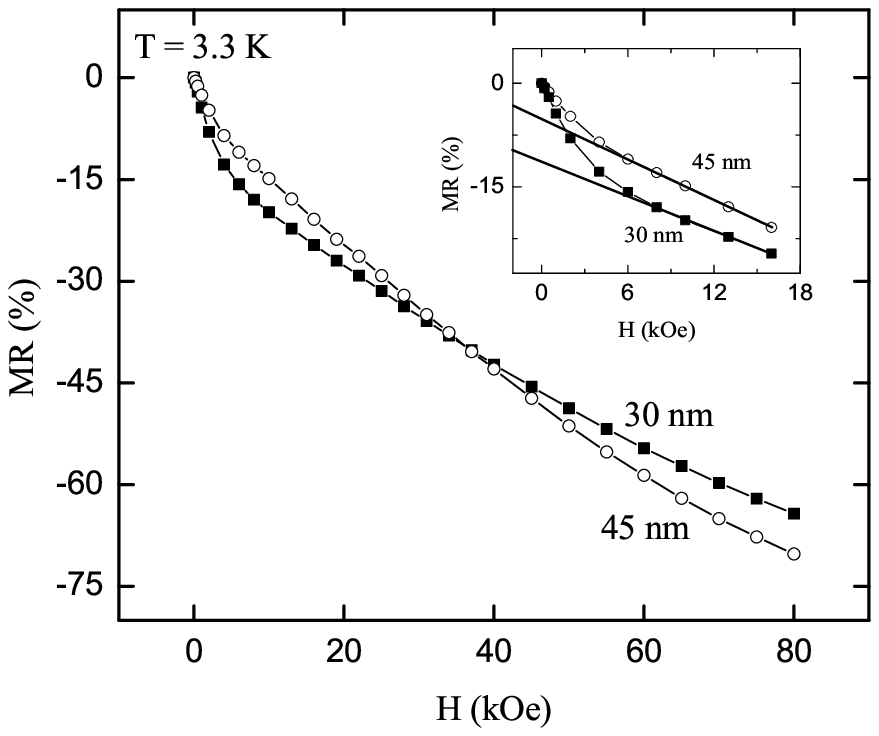}}}
\caption{}
\end{figure}

\newpage
\begin{figure}[ht]
\centering {\resizebox{8.0cm}{6.5cm}{\includegraphics{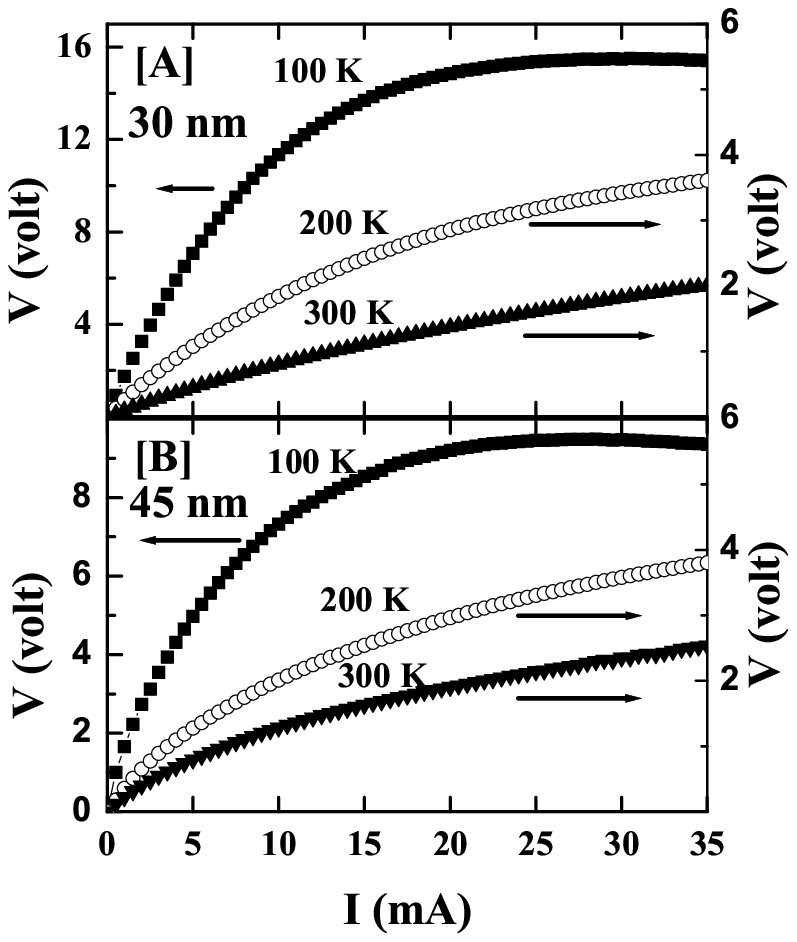}}}
\caption{}
\end{figure}
\end{document}